\documentclass[aip,apl,reprint,floatfix, 12pt]{revtex4-2}

\usepackage{microtype}
\usepackage{graphicx}
\usepackage{amsmath,amssymb,amsfonts,mathtools}
\usepackage{chemformula}
\usepackage{siunitx}
\graphicspath{{Figs/PDF/}}

\usepackage{placeins}
\newcommand*{\diameter}{\circ\kern-0.69em\diagup}
\usepackage{hyperref}
\usepackage[capitalise]{cleveref}

\usepackage{xcolor}
\usepackage{orcidlink}
\usepackage{booktabs}
\usepackage{tabularx}
\usepackage{listings}
\usepackage{dcolumn} 
\usepackage{bm} 

\usepackage[utf8]{inputenc}
\usepackage[T1]{fontenc}
\usepackage{mathptmx}
\makeatletter
\def\@email#1#2{%
 \endgroup
 \patchcmd{\titleblock@produce}
  {\frontmatter@RRAPformat}
  {\frontmatter@RRAPformat{\produce@RRAP{*#1\href{mailto:#2}{#2}}}\frontmatter@RRAPformat}
  {}{}
}%
\makeatother

\begin{document}

\title{Enhanced Thermoelectric Performance of \textit{p}-type BiSbTe Through Incorporation of Magnetic CrSb}

\author{Raphael Fortulan\orcidlink{0000-0002-1234-5212}}
\email{raphael.vicentefortulan@uwe.ac.uk}
\affiliation{Materials and Engineering Research Institute, Sheffield Hallam University, Sheffield, UK}
\affiliation{Present address: Unconventional Computing Laboratory, University of the West of England, Bristol, UK}

\author{Suwei Li}
\affiliation{School of Engineering and Material Science, Queen Mary University of London, Mile End Road, London E1 4NS, UK}

\author{Michael John Reece}
\affiliation{School of Engineering and Material Science, Queen Mary University of London, Mile End Road, London E1 4NS, UK}

\author{Illia Serhiienko}
\affiliation{International Center for Materials Nanoarchitectonics (WPI-MANA), National Institute for Materials Science, Tsukuba, Japan}
\affiliation{Graduate School of Pure and Applied Science, University of Tsukuba, Tsukuba, Japan}

\author{Takao Mori}
\affiliation{International Center for Materials Nanoarchitectonics (WPI-MANA), National Institute for Materials Science, Tsukuba, Japan}
\affiliation{Graduate School of Pure and Applied Science, University of Tsukuba, Tsukuba, Japan}

\author{Sima Aminorroya Yamini}
\affiliation{School of Aerospace, Mechanical and Mechatronic Engineering, The University of Sydney,
Sydney, 2006, Australia}

\begin{abstract}
There is evidence that magnetism can potentially increase the thermopower of materials, most likely due to magnon scattering, suggesting the incorporation of intrinsic magnetic semiconductors in non-magnetic thermoelectric materials. Here, samples of \textit{p}-type \ch{Bi_{0.5}Sb_{1.5}Te_{3}} with 10 at.\% excess Te are ball-milled with varying ratio of the antiferromagnetic semiconductor CrSb (0, 0.125, 0.5, and 1 wt.\%) to prepare bulk samples by spark plasma sintering technique. The thermopower of samples containing CrSb is increased due to an increase in the effective mass of the charge carriers, indicating that there is a drag effect originating from the magnetic particles. However, this was at the expense of reduced electrical conductivity caused by reduced charge carrier mobility. While overall only marginal improvements in power factors were observed, these samples exhibited significantly lower thermal conductivity compared to the single-phase material. As a result, a peak \textit{zT} value of $\sim$1.4 was achieved at 325 K for the sample with 0.125 wt.\% CrSb. These results highlight the potential of incorporating magnetic secondary phases to enhance the thermoelectric performance of materials.
\end{abstract}

\maketitle

	\section*{Introduction}
	To create a high-performance thermoelectric material, one must simultaneously achieve a high power factor ($\textit{PF} = \alpha^2 \sigma$) and a low total thermal conductivity ($\kappa$). However, $\alpha$, $\sigma$, and $\kappa$ exhibit strong correlations~\cite{zhangIdentifyingManipulationIndividual2021,yangUltraHighThermoelectricPerformance2020,hendricksKeynoteReviewLatest2022}, making the task of optimizing one parameter independently, while keeping the others constant, extremely challenging.
	
	Bismuth telluride alloys are among the most efficient thermoelectric materials for near-room-temperature applications. Alloying \ch{Bi2Te3} with \ch{Sb2Te3} optimizes parameters such as carrier concentration, lattice thermal conductivity, and electronic band structure to improve the figure of merit \textit{zT} = $ \nicefrac{\textit{PF}}{\kappa}T$~\cite{wittingThermoelectricPropertiesBismuth2019,wittingThermoelectricPropertiesType2020}. However, the thermoelectric performance of these alloys is limited by their relatively low thermopower.
	
	The use of magnetism has emerged as a promising strategy to increase thermopower through mechanisms such as paramagnon drag and spin-dependent effects~\cite{ahmedThermoelectricPropertiesCuGa12017,koshibaeThermopowerCobaltOxides2000,tsujiiObservationEnhancedThermopower2019,zhengParamagnonDragHigh2019}. The concept that magnetism can possibly enhance thermoelectric properties dates back decades, with early work proposing that magnon scattering could be the origin of an increase in the thermopower of some magnetic elements at low temperatures~\cite{bailynMaximumVariationalPrinciple1962}. These include (1) using intrinsically magnetic semiconductors, such as MnTe~\cite{wasscherContributionMagnondragThermoelectric1964}, \ch{CuFeS2}~\cite{tsujiiHighThermoelectricPower2013,angThermoelectricityGenerationElectronMagnon2015}, and \ch{Cr2Ge2Te6}~\cite{pengImprovementThermoelectricityMagnetic2018}; (2) doping non-magnetic thermoelectric materials with magnetic elements resulting in increased thermopower due to interactions between charge carriers and local magnetic moments ~\cite{ahmedThermoelectricPropertiesCuGa12017, vaneyMagnetismmediatedThermoelectricPerformance2019,acharyaCouplingChargeCarriers2018,dasThermoelectricPropertiesMn2019,matsuuraTheoryHugeThermoelectric2021}; and (3) introducing magnetic secondary phases, such as nanoparticles or inclusions, into non-magnetic thermoelectric matrices~\cite{zhaoMagnetoelectricInteractionTransport2017, luCoherentMagneticNanoinclusions2019, liuThermoelectricPropertiesBulk2011,bourgesInvestigationMnSingle2023}. Studies on \ch{Ba_{0.3}In_{0.3}Co4Sb12}~\cite{gaoCriteriaPowerFactor2018, zhaoMagnetoelectricInteractionTransport2017} and \ch{Ti_{0.25}Zr_{0.25}Hf_{0.5}NiSnSb}~\cite{luCoherentMagneticNanoinclusions2019} materials have shown that the inclusion of coherent magnetic particles can simultaneously enhance thermopower and carrier mobility. Magnetic secondary phases may allow the tuning of properties through the composition, size, and microstructure of the materials~\cite{liuCandidateMagneticDoping2019, marghussianMagneticPropertiesNanoGlass2015, tanValenceBandModification2015, vandendriesscheMagnetoopticalHarmonicSusceptometry2013, zhaoSuperparamagneticEnhancementThermoelectric2017}.
 
	In this study, ball-milled stoichiometric \ch{Bi_{0.5}Sb_{1.5}Te{3}} 10 at.\% Te-rich and CrSb (0, 0.125, 0.5, and 1 wt.\%) samples were fabricated by spark plasma sintering (SPS). Excess Te was added to the system as Te-rich bismuth telluride and bismuth antimony telluride alloys showed high thermoelectric performance~\cite{zhuangThermoelectricPerformanceEnhancement2021,liuCrystallographicallyTexturedNanomaterials2018,dengThermalConductivityBi02018,chauhanCompositionalFluctuationsMediated2021}. The inclusion of excess Te suppresses the defects caused by its easy volatilization during the SPS process~\cite{kimHighThermoelectricPerformance2017} and this strategy has been shown to result in more efficient thermoelectric materials~\cite{chauhanCompositionalFluctuationsMediated2021}. CrSb is an antiferromagnetic semiconductor with a N\'eel temperature of approximately \SI{680}{\kelvin}~\cite{dohnomaeMagneticStructureCrSb1993}, which makes it a promising magnetic secondary phase candidate for improving the thermoelectric performance for room temperature applications. By incorporating varying concentrations of magnetic CrSb particles, we were able to increase the thermopower through indicated drag effect, while maintaining a relatively high electrical conductivity. The addition of a secondary phase introduces additional phonon scattering mechanisms that reduce thermal conductivity. This reduction combined with a higher thermopower, enabled a high peak of \textit{zT} $\approx 1.4$ at \SI{325}{\kelvin}.
	
	\section*{Experimental details}
	\subsection*{Synthesis}
	Polycrystalline \ch{Bi_{0.5}Sb_{1.5}Te_{3+0.3}} samples were synthesized by direct reaction of stoichiometric amounts of Bi (99.999\%, Alfa Aesar), Te (99.999\%, Alfa Aesar), and Sb (99.999\%, Alfa Aesar) in vacuum-sealed quartz ampules in an inert atmosphere glovebox. The ampules were heated to \SI{850}{\celsius} for \SI{12}{\hour}, homogenized at \SI{1000}{\celsius}, quenched in cold water, and annealed at \SI{400}{\celsius} for \SI{72}{\hour}.

    A pristine CrSb sample was synthesized by loading stoichiometric amounts of Cr (99.95\%, Alfa Aesar) and Sb (99.999\%, Alfa Aesar) into a vacuum-sealed quartz ampule. The ampule was heated to \SI{850}{\celsius} for \SI{24}{\hour}, mixed every \SI{4}{\hour}, homogenized at \SI{1160}{\celsius} for \SI{1}{\hour}, and allowed to cool naturally. The resulting ingot was hand-ground using an agate mortar and pestle in a glovebox, loaded into a vacuum-sealed quartz ampule, and annealed at \SI{900}{\celsius} for \SI{24}{\hour}. The annealed powder was then sintered in a graphite die under vacuum to make $\diameter$ \SI{11}{\milli\meter} rods using SPS (KCE FCT-H HP D-25 SD, FCT Systeme GmbH, Rauenstein, Germany) at \SI{50}{\mega\pascal} for \SI{20}{\minute} at \SI{900}{\celsius}.
    
    Samples of \ch{Bi_{0.5}Sb_{1.5}Te_{3}} with $x$ wt.\% CrSb ($x$ = 0, 0.125, 0.5, and 1) were fabricated by wet ball milling powdered ingots of \ch{Bi_{0.5}Sb_{1.5}Te_{3+0.3}} and CrSb. The cast ingots were pre-milled using an agate mortar and pestle in a glovebox. The powders were weighed and placed in a \SI{250}{\milli\litre} agate jar with \SI{20}{\milli\meter} agate balls and ethanol (99.97\%, VWR). The ball-to-powder ratio was 15:1, and the solvent-to-powder ratio was \SI{100}{\milli\litre} to \SI{10}{\gram}~\cite{kanatziaDesignBallMillingExperiments2013}. Milling was performed using a Retsch Planetary Ball Mill PM 100 at 300 rpm for \SI{4}{\hour} at \SI{15}{\minute} intervals with \SI{5}{\minute} breaks and a change in direction halfway through. The jar was then placed in a desiccator for at least \SI{15}{\hour}. The dried powders were sintered in a graphite die under vacuum to make $\diameter$ \SI{11}{\milli\meter} rods using SPS at \SI{50}{\mega\pascal} for \SI{5}{\minute} at \SI{400}{\celsius}. The densities of all samples were approximately 95\% of their nominal density.

	\subsection*{Materials characterization}
	The phase purity and crystal structure of the sintered samples were characterized by Powder X-ray Diffraction (PXRD) using a PANalytical X'Pert Pro diffractometer with CuK$\alpha$1 radiation ($\lambda = \SI{0.15406}{\nano\metre}$, \SI{40}{\kilo\volt}, \SI{40}{\milli\ampere}).
	
	Electrical conductivity ($\sigma$) and thermopower ($\alpha$) were measured perpendicular to the sintering direction of the samples by cutting $\approx$$2 \times 2 \times 8$ \si{\cubic\milli\metre} bar specimens from the rods. Measurements were conducted from room temperature to 523 K under a helium atmosphere using a Linseis LSR-3 apparatus.
	
	The thermal diffusivity ($D$) of all the samples was measured by the LFA method using a NETZSCH LFA 467 HyperFlash\textsuperscript{\textregistered} instrument. Slab-shaped samples were also cut to measure the room-temperature Hall coefficient ($R_\mathrm{H}$) under a $\pm$\SI{0.55}{\tesla} magnetic field using an ECOPIA 3000 Hall Effect Measurement System. The Hall carrier concentration ($n_\mathrm{H}$) was calculated as $n_\mathrm{H} = 1 / (e\cdot R_\mathrm{H})$.
	
	Heat capacity measurements were conducted from room temperature to 473 K using a PerkinElmer DSC 8000 according to the ASTM sapphire standard method E1269~\cite{e37committeeTestMethodDetermining}. The measured heat capacity values are presented in the Appendix~\ref{sec:heat_capacity}.
	\section*{Results and discussion}
	\subsection*{Structural and phase analysis}
\begin{figure}[hb]
		\centering
		\includegraphics[width=1\linewidth, keepaspectratio]{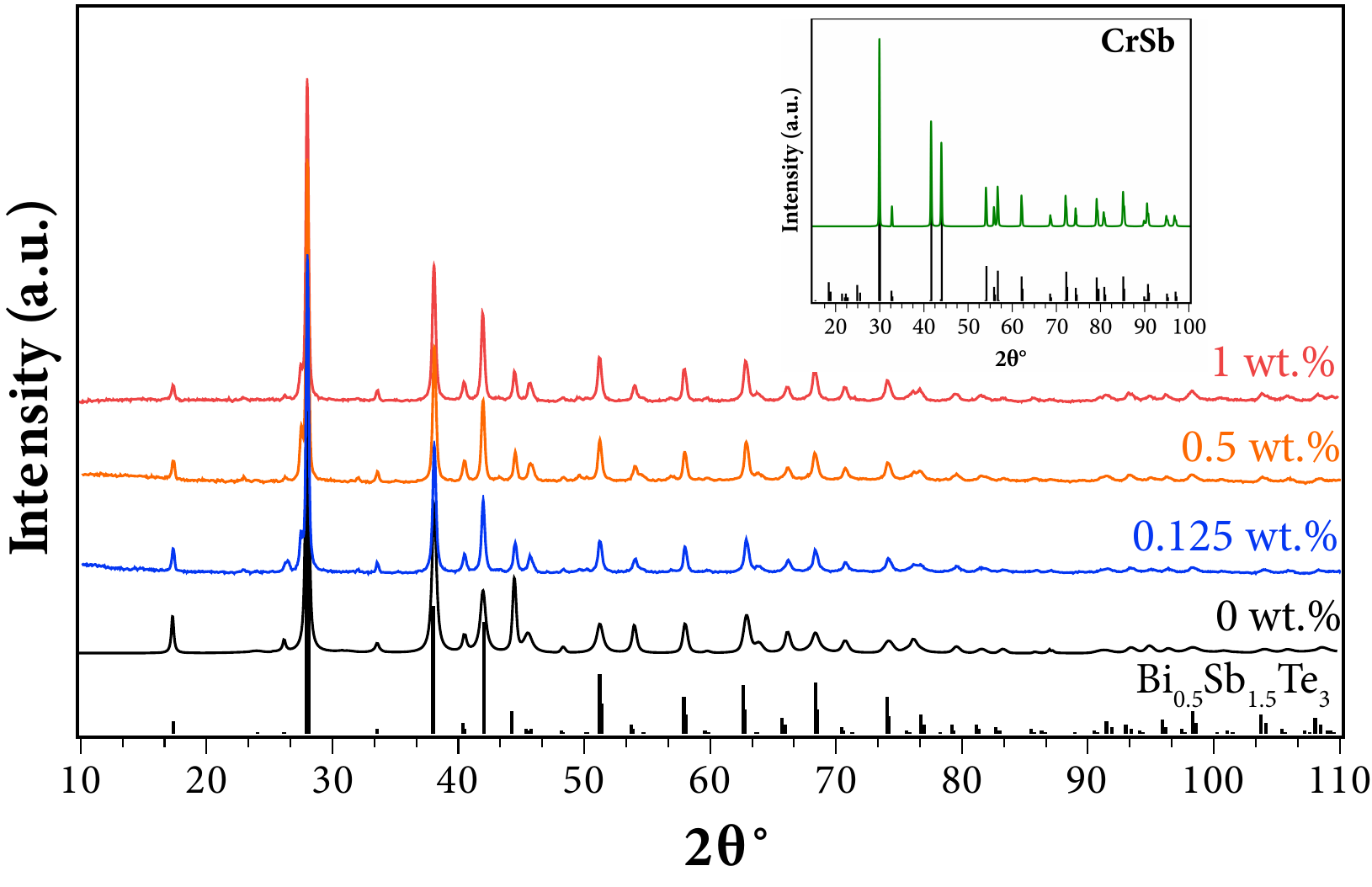}
		\caption{Powder X-ray diffraction patterns of \ch{Bi_{0.5}Sb_{1.5}Te_{3 +0.3}} with $x$ wt.\% CrSb ($x$ = 0, 0.125, 0.5, and 1) samples. The inset shows the diffraction pattern for pristine CrSb.}
		\label{fig:bisbte_crsb_ball_XRD}
	\end{figure}

    The PXRD patterns of all samples in \cref{fig:bisbte_crsb_ball_XRD} match the rhombohedral \ch{Bi_{0.5}Sb_{1.5}Te_{3}} phase. The inset shows that the diffraction pattern of the pristine CrSb sample corresponds to a single-phase CrSb hexagonal structure.
	\subsection*{Transport properties}
\begin{figure*}[!ht]
		\centering
        \includegraphics[width=\textwidth,keepaspectratio]{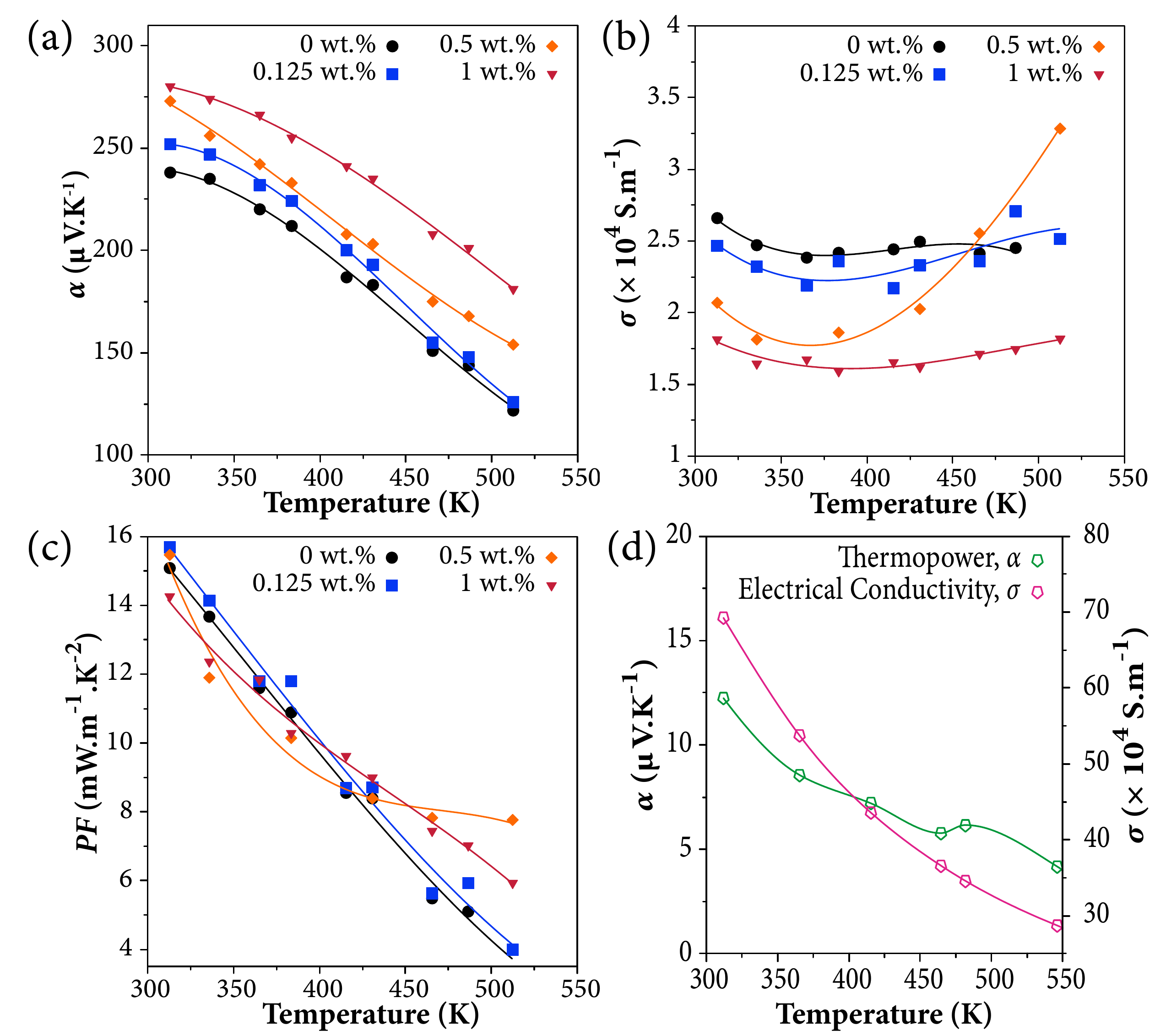}
		\caption{Temperature dependence of the (a) thermopower, (b) electrical conductivity, and (c) power factor of \ch{Bi_{0.5}Sb_{1.5}Te_{3 +0.3}} with $x$ wt.\% CrSb ($x$ = 0, 0.125, 0.5, and 1) samples, and (d) thermopower and electrical conductivity of pristine CrSb.}
		\label{fig:transport_bisbte_crsb_ball}
	\end{figure*}

	Figs.~\ref{fig:transport_bisbte_crsb_ball}(a) to (c) show the thermopower, electrical conductivity, and power factor of \ch{Bi_{0.5}Sb_{1.5}Te_{3 +0.3}} with $x$ wt.\% CrSb ($x$ = 0, 0.125, 0.5, and 1), measured perpendicular to the direction of sintering. All samples showed a positive thermopower (\cref{fig:transport_bisbte_crsb_ball}(a)), indicating \textit{p}-type semiconductor behavior. The electrical conductivity exhibited metallic behavior with decreasing values, as seen in \cref{fig:transport_bisbte_crsb_ball}(b).
 
	The room-temperature values of the thermopower increase with the concentration of the CrSb, rising from $\sim$\SI{238}{\micro\volt\per\kelvin} for the single-phase sample to $\sim$\SI{280}{\micro\volt\per\kelvin} for the 1 wt.\% CrSb. With the opposite effect on electrical conductivity, the electrical conductivity of the single-phase sample decreased from $\sim$\SI{2.7e4}{\siemens\per\meter} for the single-phase sample to $\sim$\SI{1.8e4}{\siemens\per\meter} for the sample with 1 wt.\% CrSb.
	
	\cref{fig:transport_bisbte_crsb_ball}(d) shows the temperature-dependent electrical transport properties of pristine CrSb samples measured perpendicularly to the sintering direction. The electrical conductivity and thermopower are comparable to the one seen in the literature~\cite{polashMagnondragThermopowerAntiferromagnets2020}.

	\cref{tab:hall_ball_milled} lists the Hall carrier concentrations and mobilities of the samples. The carrier concentration of these samples is lower than those observed in the literature ($\sim$\SI{1e19}{\per\cubic\centi\metre}~\cite{chauhanCompositionalFluctuationsMediated2021}) which results in the overall lower electrical conductivity of these samples. The mobility of \SI{232}{\centi\meter\squared\per\volt\per\second} for the single-phase samples is similar to those reported in the literature (e.g., \SI{248}{\centi\meter\squared\per\volt\per\second}~\cite{zhuangThermoelectricPerformanceEnhancement2021}). The values of $n_\mathrm{H}$ ranged from $7.7\sim 7.1$ \SI{e18}{\per\cubic\centi\metre}, indicating that the excess Te stabilized the carrier concentration of the samples. However, the mobility decreases with the inclusion of the secondary phase, which is consistent with the behavior of multiphase materials ~\cite{koEnhancedThermopowerCarrier2011,zhangExcellentThermoelectricPerformance2021}.
	\begin{table}[ht]
		\centering
		\caption{Room temperature Hall carrier concentration and mobility of ball milled \ch{Bi_{0.5}Sb_{1.5}Te_{3 + 0.3}} with $x$ wt.\% CrSb ($x$ = 0, 0.125, 0.5, and 1) samples}
		\label{tab:hall_ball_milled}
		\begin{tabular}{lrr}
			\toprule
			$x$ & {$n_\mathrm{H} (\times$ \SI{e18}{\per\cubic\centi\metre})} & $\mu_\mathrm{H}$ (\si{\centi\meter\squared\per\volt\per\second}) \\ \midrule
			0 & 7.7 & 232.0 \\
			0.125 & 7.7 & 203.9 \\
			0.5 & 7.3 & 185.5 \\
			1 & 7.1 & 161.6 \\ \bottomrule
		\end{tabular}
	\end{table}

    	\begin{figure*}[!t]%
        \centering
        \includegraphics[width=1\textwidth,keepaspectratio]{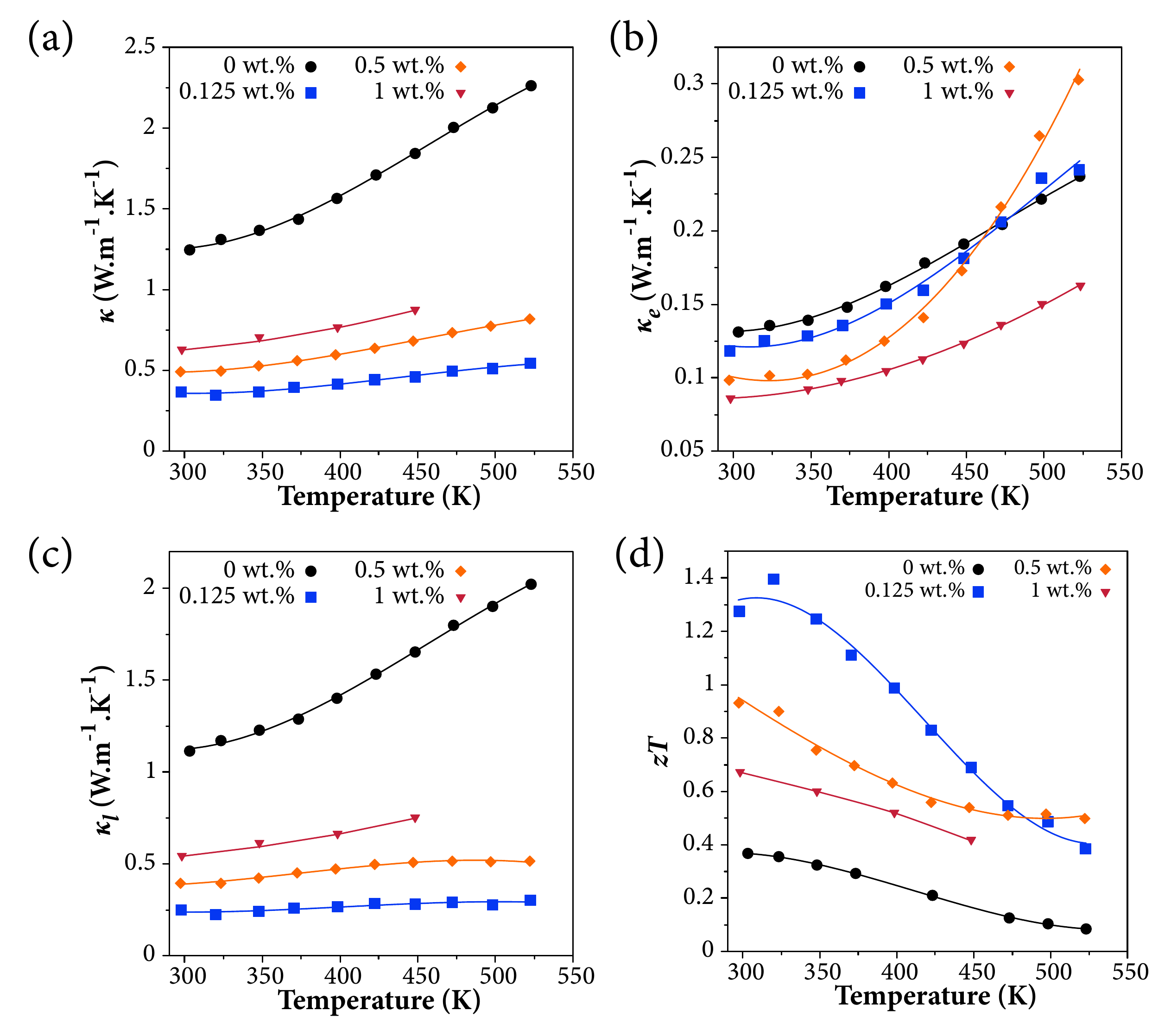}
		\caption{Temperature dependence of the (a) thermal conductivity, (b) electronic thermal conductivity, (c) lattice thermal conductivity, (d) figure of merit \textit{zT} of \ch{Bi_{0.5}Sb_{1.5}Te_{3+0.3}} with $x$ wt.\% CrSb ($x$ = 0, 0.125, 0.5, and 1) samples.}
		\label{fig:transport_bisbte_crsb_2}
	\end{figure*}
 
	Since the carrier concentration of all samples has similar values, these changes cannot simply be attributed to changes in the values of $n_\mathrm{H}$. The behavior of the band structure with CrSb inclusion was analyzed by modeling the thermopower and Hall carrier concentration using the single parabolic band (SPB) model~\cite{snyderComplexThermoelectricMaterials2008} (discussed in the Appendix~\ref{sec:spb}). The SPB model used here may not fully and accurately determine the behavior of multiphase samples due to the presence of bipolar conduction, complex scattering processes, and the non-parabolic nature of the valence band. However, this approach may indicate a trend in the calculated effective mass, $m^*$, of samples~\cite{zhengMechanicallyRobustBiSbTe2015}.
	\begin{table}[!ht]
		\centering
		\caption{Calculated effective mass of ball milled \ch{Bi_{0.5}Sb_{1.5}Te_{3 +0.3}} with $x$ wt.\% CrSb ($x$ = 0, 0.125, 0.5, and 1) samples using the single parabolic band model}
		\label{tab:pisarenko_ball_milled}
		\begin{tabular}{lr}
			\toprule
			$x$   & $m^* (m_0)$ \\ \midrule
			0     & 0.90 \\
			0.125 & 1.16 \\
			0.5   & 1.06 \\
			1     & 1.21 \\ \bottomrule
		\end{tabular}
	\end{table}
 
 	The effective mass of samples is increased in the presence of CrSb \cref{tab:pisarenko_ball_milled}. Since the thermopower for a degenerate semiconductor, with a parabolic band and energy-independent scattering approximation, can be written as~\cite{snyderComplexThermoelectricMaterials2008}
	\begin{equation}
		\alpha = \frac{8 \pi^2 k_{\mathrm{B}}}{3q h^2} m^{*}T\left(\frac{\pi}{3n}\right)^{2/3},
		\label{eq:seebeck_parabolic_band}
	\end{equation}
	where $m^*$ is the effective mass, \cref{eq:seebeck_parabolic_band}  suggests that the presence of a magnetic phase in the material can lead to an increase in the effective mass of the sample and consequently an increase in the thermopower, similar to observations of studies with magnetic dopants~\cite{fortulanThermoelectricPerformanceNType2022,vaneyMagnetismmediatedThermoelectricPerformance2019}.


    The results show that despite the presence of a magnetic secondary phase with poor thermoelectric performance and low thermopower of approximately \SI{12.5}{\micro\volt\per\kelvin} at room temperature and \SI{4.38}{\micro\volt\per\kelvin} at \SI{550}{\kelvin}, the electronic performance of the samples can be improved rather than degraded (see \cref{fig:transport_bisbte_crsb_ball}(d)).
	
	The combined decrease in carrier mobility due to the presence of an additional phase and the possible dragging effect caused by the magnetic phase degrade the electrical conductivity of multiphase samples~\cite{vaneyMagnetismmediatedThermoelectricPerformance2019}. However, the overall effect, was an increase in the power factor $PF$ was detected (as shown in \cref{fig:transport_bisbte_crsb_ball}(c)).

    The thermal conductivity ($\kappa$) of the samples is shown in \cref{fig:transport_bisbte_crsb_2}(a). The electronic contribution to the thermal conductivity ($\kappa_e$) is shown in \cref{fig:transport_bisbte_crsb_2}(b) and it was estimated using the Wiedemann-Franz law~\cite{jonsonMottFormulaThermopower1980} ($\kappa_e=L\sigma T$) where $L$ is the Lorenz number, and it was calculated using the SPB model (see Appendix~\ref{sec:spb} for more details).
    	
    The lattice ($\kappa_l$) contributions to thermal conductivity were calculated as $\kappa_l = \kappa - \kappa_e$, as shown in \cref{fig:transport_bisbte_crsb_2}(c). The measured values for the lattice thermal conductivity are optimally low reaching values slightly below the glass-like thermal conductivity for \ch{Bi2Te3} ($\kappa_{\mathrm{glass}}\approx \SI{0.31}{\watt\per\metre\per\kelvin}$~\cite{liuCrystallographicallyTexturedNanomaterials2018}) but higher than Cl-doped \ch{Bi2Te3} alloys (\SI{0.15}{\watt\per\metre\per\kelvin}~\cite{parashchukUltralowLatticeThermal2022}), namely the sample containing the 0.125 wt.\% of CrSb reached the lowest value of $\sim \SI{0.24}{\watt\per\metre\per\kelvin}$. The low values for all samples indicate a high degree of scattering of the phonons due to additional boundaries introduced by the secondary phase.
    \FloatBarrier
     The figure of merit (\textit{zT}) of the samples is shown in \cref{fig:transport_bisbte_crsb_2}(d). The combination of significantly reduced thermal conductivity, aligned with an increase in thermal power due to the incorporation of the magnetic secondary phase, contributed to the high \textit{zT} values for the multiphase samples compared to the single-phase \ch{Bi_{0.5}Sb_{1.5}Te_{3 +0.3}} material. We note that our single-phase sample has a relatively low \textit{zT} value of around 0.36, similar to what was observed for~\cite{weiSignificantlyEnhancedThermoelectric2017}, and indicate the maximum \textit{zT} for this sample was below room temperature~\cite{banoEnhancedThermoelectricPerformance2021}. 
    
    The 0.125 wt.\% CrSb sample exhibited the highest \textit{zT}, reaching a peak value of $\sim$1.4 at 325 K. The other multiphase compositions also showed promising \textit{zT} improvements, albeit to a lesser extent. The 0.5 wt.\% CrSb had the next highest \textit{zT} of $\sim$0.92 at 525 K, followed by the 1 wt.\% CrSb at $\sim$0.68. The high \textit{zT} values result from the synergistic effects of magnetically induced thermopower enhancement and thermal conductivity reduction due to interfacial and magnetic scattering in the multiphase samples.
    
\begin{figure}[!ht]
    \centering
        \includegraphics[width=1\linewidth, keepaspectratio]{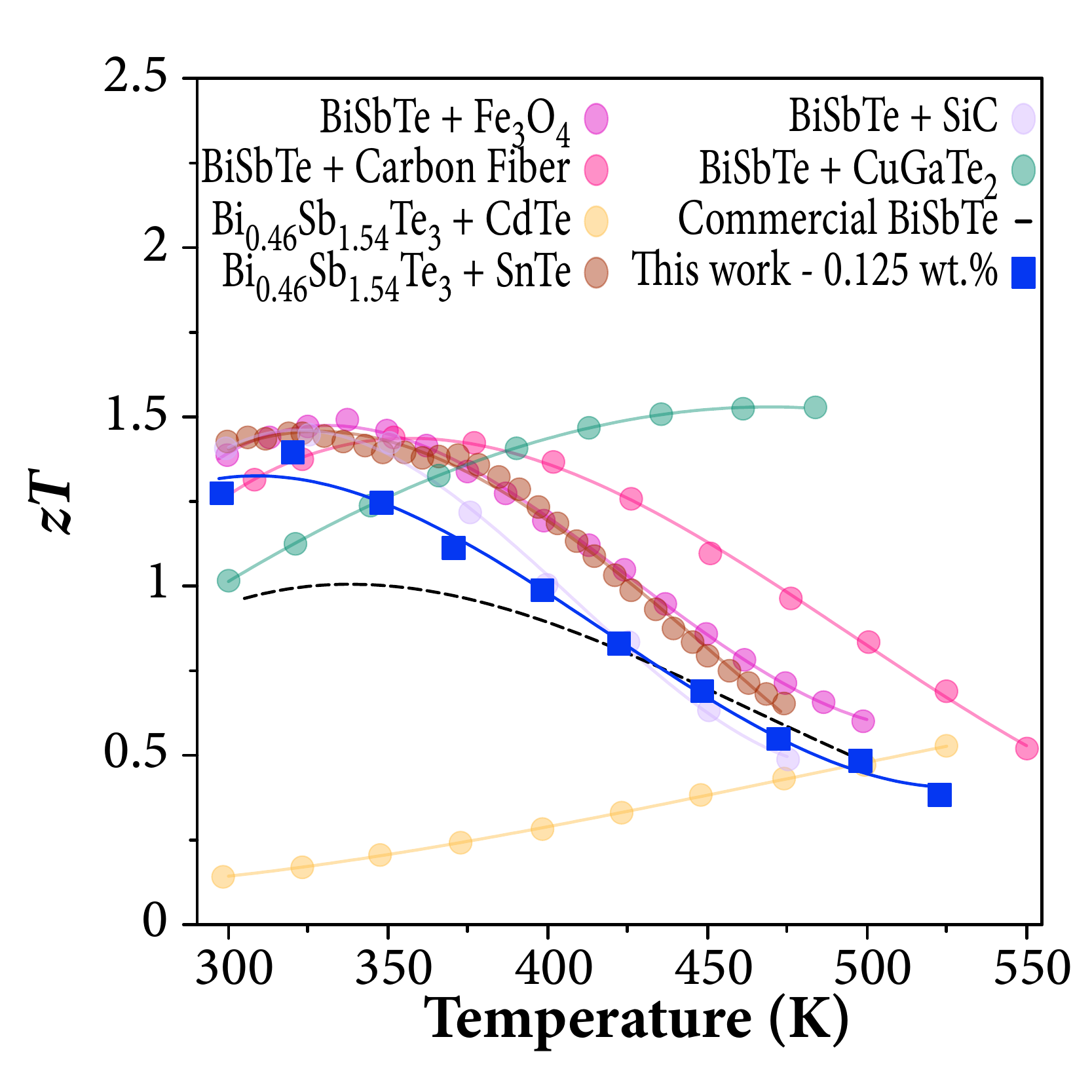}
        \caption{Temperature dependence of \textit{zT} of \ch{Bi_{0.5}Sb_{1.5}Te_{3 +0.3}} with $x$ wt.\% CrSb ($x$ = 0, 0.125, 0.5, and 1) samples. Comparisons of \textit{zT} values of \ch{Bi_{0.5}Sb_{1.5}Te_{3 +0.3}} + 0.125 wt.\%, commercial BiSbTe, and other typical \textit{p}-type BiSbTe composites: \ch{Bi_{0.5}Sb_{1.5}Te_3} + \ch{Fe3O4}~\cite{liMagnetisminducedHugeEnhancement2020}, \ch{Bi_{0.5}Sb_{1.5}Te_3} + carbon fiber~\cite{yangSignificantEnhancementThermoelectric2021}, \ch{Bi_{0.46}Sb_{1.54}Te_3} + CdTe~\cite{taoEnhancedThermoelectricPerformance2020}, \ch{Bi_{0.46}Sb_{1.54}Te_3} + SnTe~\cite{zhangHighThermoelectricPerformance2019}, \ch{Bi_{0.46}Sb_{1.54}Te_3} + SiC~\cite{zhangEnhancedThermoelectricPerformance2019}, and \ch{Bi_{0.4}Sb_{1.6}Te3} + \ch{CuGaTe2}~\cite{liHighThermoelectricEfficiency2019}.}
        \label{fig:zT}
    \end{figure}

    In \cref{fig:zT}, the results obtained in this work are compared with other compositions seen in the literature~\cite{liMagnetisminducedHugeEnhancement2020,yangSignificantEnhancementThermoelectric2021,taoEnhancedThermoelectricPerformance2020,zhangHighThermoelectricPerformance2019,zhangEnhancedThermoelectricPerformance2019,liHighThermoelectricEfficiency2019}, as well as with a commercial BiSbTe ingot. The introduction of the CrSb phase in the \ch{Bi_{0.5}Sb_{1.5}Te_{3 +0.3}} material shows a performance that surpasses the one seen for the commercial ingot and is comparable to the best results reported in the literature for composite BiSbTe materials.
    \FloatBarrier
    \section*{Conclusions}
    A series of \ch{Bi_{0.5}Sb_{1.5}Te_{3 +0.3}} samples with varying concentrations of CrSb magnetic secondary phase (0, 0.125, 0.5, and 1 wt.\%) were synthesized by a combination of ball milling and spark plasma sintering techniques. The results showed that the incorporation of small amounts of the CrSb magnetic phase significantly enhanced the thermopower of the samples by increasing the carriers’ effective mass, which is consistent with previous findings for magnetic dopants. However, the electrical conductivity is adversely affected by the reduced carrier mobility caused by the presence of the secondary phase. The increased in the powe factor, combined with the significantly lower thermal conductivity resulted in a high figure of merit (\textit{zT}) values for CrSb added samples. These results confirm the potential benefits of incorporating magnetic secondary phases into thermoelectric materials to modulate their electronic and thermal transport properties favorably.
\section*{Author Declarations}
    \subsection*{Data availability}
    The authors declare that the data supporting the findings of this study are available within the paper. Should any raw data files be needed in another format they are available from the corresponding author upon reasonable request. 
    \subsection*{Declaration of competing interest}
    The authors have no conflicts to disclose.
    \begin{acknowledgments}
    This study was supported by the European Union’s Horizon 2020 research and innovation program under the Marie Sklodowska-Curie Grant Agreement No. 801604. This work also received support from the Henry Royce Institute for Advanced Materials, funded through EPSRC grants EP/R00661X/1, EP/S019367/1, EP/P025021/1, and EP/P025498/1. TM would like to thank JST Mirai Program Grant Number JPMJMI19A1. IS was supported by JST SPRING, Grant Number JPMJSP2124.
    \end{acknowledgments}
    
\appendix
    \section{Heat capacity measurements}
    \label{sec:heat_capacity}
\begin{figure}[!hb]%
	\centering
	\includegraphics[width=.5\textwidth, keepaspectratio]{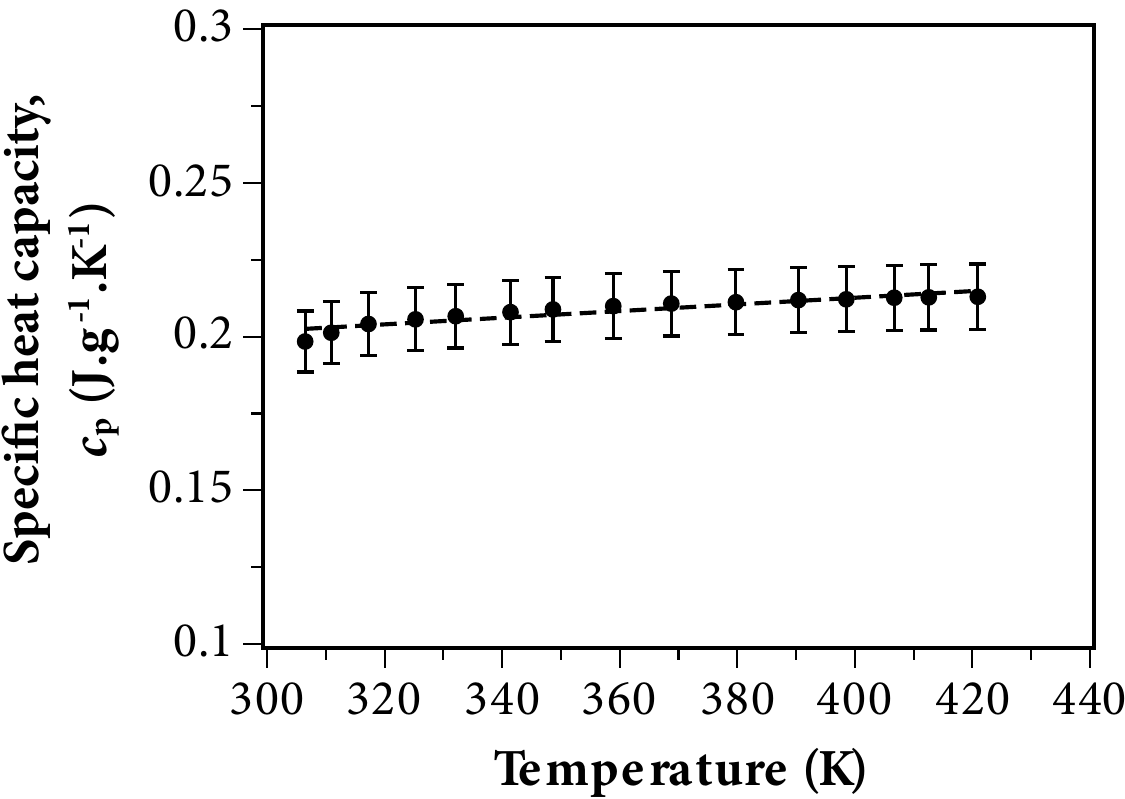}
	\caption{Temperature-dependent specific heat capacity of \ch{Bi_{0.5}Sb_{1.5}Te_{3+0.3}}. The dashed line represents a linear fit of the data.}
	\label{fig:heat_capacity_bisbte_crsb}%
\end{figure}

    The heat capacity of \ch{Bi_{0.5}Sb_{1.5}Te_{3+0.3}} is displayed in~\cref{fig:heat_capacity_bisbte_crsb}.
\section{Single parabolic band model}
    \label{sec:spb}
    For the single parabolic band (SPB) model with acoustic phonons as the main scattering mechanisms, the thermopower, carrier concentration, and Lorenz number are evaluated as
    \begin{subequations}
        \begin{align}
            & \alpha =\pm \frac{k_\mathrm{B}}{e}\left(\frac{2 F_{1}(\eta)}{F_{0}(\eta)}-\eta\right), \label{eq:Seebeck_SPB}\\
            & n =  \frac{\left(2 m^{*} k_\mathrm{B} T \right)^{3/2}}{2\pi^2\hbar^3}F_{1/2}\left(\eta\right),\label{eq:carrier_SPB}\\
            & L = \left(\frac{k_\mathrm{B}}{e}\right)^2 \frac{3F_0(\eta)F_{2}(\eta)-4 F_{1}(\eta)^2}{F_{0}(\eta)^2},\label{eq:L_SPB}
        \end{align}
    \end{subequations}
    where $e$ is the elementary charge, $k_\mathrm{B}$ is the Boltzmann constant, $\varepsilon = \frac{E}{k_\mathrm{B} T}$ is the reduced energy, $\eta = \frac{E_F}{k_\mathrm{B} T}$ is the reduced Fermi level, $\hbar$ is the reduced Planck constant, $m^{*}$ is the density of states effective mass.
    
    $F_j(\eta)$ is the Fermi-Dirac integral for an index $j$ and is defined as
    \begin{equation}
        F_j(\eta) \coloneqq \int_{0}^{\infty} \frac{\varepsilon^j}{\exp(\varepsilon - \eta) + 1}\mathrm{d}\varepsilon.
    \end{equation}
    
    Since \cref{eq:Seebeck_SPB} is independent of \cref{eq:carrier_SPB}, the procedure for estimating the effective mass was to first calculate the reduced Fermi level from the thermopower measurement and then estimate $m^{*}$ from the measured carrier concentration.
\FloatBarrier
\section*{References}
\bibliography{References_V2}
\end{document}